# Unexpected bending behavior of 2D lattice materials

Yu Yao, Yong Ni, and Ling-hui He*

*CAS Key Laboratory of Mechanical Behavior and Design of Materials, University of Science and Technology of China, Hefei, Anhui 230026, China*

## Abstract

Architected 2D lattice materials are appealing for shape-shifting applications due to the tunable sign of Poisson's ratio. It is commonly believed that the positive and negative Poisson's ratios lead to anticlastic and synclastic curvatures respectively when the material is bent in one direction. Here, taking 2D beam lattices with star-shaped unit cells as examples, we show theoretically and demonstrate experimentally that this is not always true. At a fixed Poisson's ratio, we find a transition between anticlastic and synclastic bending curvatures controlled by the beam's cross-sectional aspect ratio. Such an unexpected behavior roots in the competition between torsion and bending of the beams, and can be well captured by a Cosserat continuum model.

## 1. Introduction

Mechanical metamaterials with engineered structures are of growing interest for they enable extraordinary deformations unaccessible by conventional materials (1-6). A typical kind of such materials is architected 2D beam lattices which exhibit either positive or negative in-plane Poisson's ratios (IPRs) depending on unit cell pattern (3, 7-10). Because a conventional elastic plate with positive IPR reveals anticlastic curvature (saddle shape) when bent in one direction (11), it has been commonly believed as a rule that 2D lattices with positive IPR also behave so while those with negative IPR adopt synclastic curvature (dome shape) in the same loading condition (8, 12-24). Indeed, both kinds of curvatures are observable in conventional and re-entrant 2D hexagonal honeycombs which possess positive and negative IPRs respectively (12).

---

* Corresponding author: lhhe@ustc.edu.cn

The phenomenon of synclastic bending has inspired innovative ways in shape-shifting applications, such as manufacturing of doubly curved aircraft structures(14), construction of bending-active architectures (24), and design of adaptive fashion, furniture, sport equipment (21), as well as soft wearable robotics (25).

Much effort has been done to understand in-plane deformation of the architected 2D materials; for the large number of literature we mention some recent studies (2, 3, 20, 26-35) among others. In contrast, analysis of the out-of-plane bending is rare (14, 18, 23, 36-42). Worth mentioning is the finite element modeling (FEM) (18) which predicted that a 2D anti-trichiral cylinder-ligament honeycomb possessing negative IPR exhibits anticlastic bending shape in short ligament limit. This does not mean an exception, because in that case the cylinder rotation mechanism is redundant so that the material actually bends like a hexagonal honeycomb with positive IPR. In fact, no conflicts with the recognized bending rule have been reported so far, but the underlying microstructural origin of the curvature remains elusive.

Here, we combine theory and experiments to explore unidirectional bending of 2D lattice materials with star-shaped unit cells consisting of elastic beams of rectangular cross-section. Though there are other structural forms, the star-shaped construction offers a convenient way to alter the sign of IPR of the material by changing the re-entrant angle. Surprisingly, we find that the bending curvature is convertible between anticlastic and synclastic by changing the beam's cross-sectional aspect ratio while keeping the IPR of the material invariant. In other words, whatever the sign of the IPR, the lattice material can bend into saddle, cylinder, or dome shapes at different thicknesses. Our result defies the common belief, and is attributed to the competition between axial torsion and out-of-plane bending of the beams. A continuum model based on Cosserat elasticity (43, 44) is developed to capture the unusual bending behavior.

## 2. Results and Discussion

***Tetragonal 2D lattices***. We begin with examining a tetragonal 2D lattice material sketched in Fig. 1*A*, where a rectangular coordinate system $(x_1, x_2, x_3)$ is introduced with $x_3 = 0$ lying in the mid-plane. For simplicity, the length of the connecting struts is assumed to be twice the side length $l$ of the four-pointed stars. The width and height of the cross-section of the beams are $b$ and $h$, respectively, and the re-entrant angle of the stars is $2\theta$. To analyze the uniform deformation of the material, we focus on a four-node square unit cell shown in Fig. 1*B*, which has different variants if $\theta$ varies in the range $\theta \in (0^\circ, 180^\circ)$ (Fig. 1C). We suppose that the beams are slender enough so that the Euler-Bernoulli beam theory (45) can be applied.

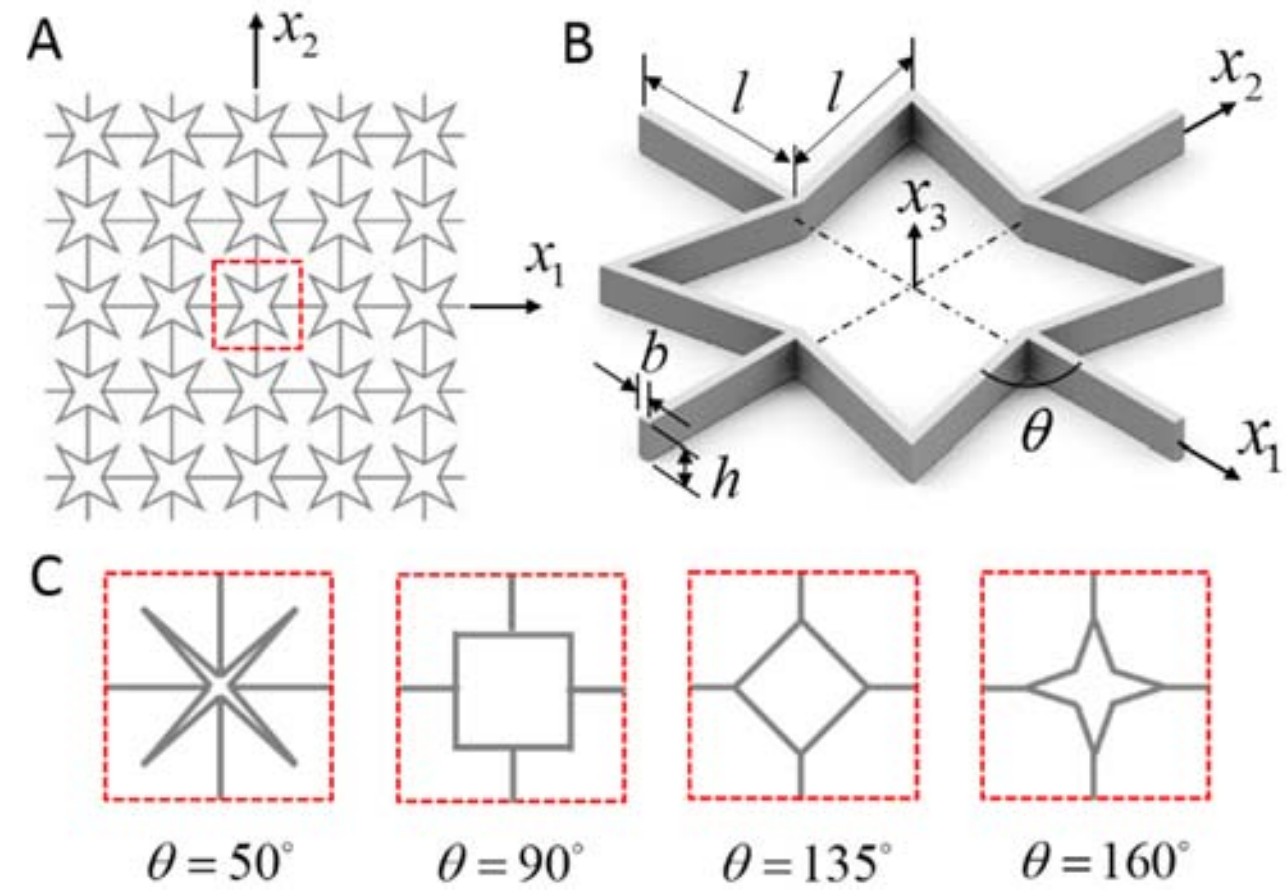

FIG. 1. (*A*) Schematic of the structure of the tetragonal 2D lattice material. (*B*) Geometric detail of the 4-node square unit cell, where the $x_1 - x_2$ plane coincides with the undeformed mid-plane. (*C*) Variants of the unit cell formed at different re-entrant angles of the star.

The equivalent IPR $\nu_s(\phi)$ of the material depends on the loading direction $\phi$ with respect to the $x_1$ axis. If tensioned uniformly along the $x_1$ axis ($\phi = 0^\circ$), the deformation in the lattice exhibits mirror symmetry about the $x_1 - x_3$ and $x_2 - x_3$ planes. That is, the struts parallel to the $x_1$ axis elongate, the sides of the stars are tensioned axially and bent in the plane, and the struts parallel to the $x_2$ axis do not deform but experience in-plane shift. The material remains planar. In small deformations, one can readily derive the relative displacements between two opposite nodes in the unit cell, yielding the average strains $\varepsilon_{11}$ and $\varepsilon_{22}$ in the $x_1$ and $x_2$ directions respectively. The equivalent IPR $\nu_s(0) = -\varepsilon_{22} / \varepsilon_{11}$ is obtained as (*SI Appendix*, S1A)

$$\nu_s(0) = \frac{3-[5-2(b/l)^2]\sin 2\theta}{5-3\sin 2\theta + 6(b/l)^2}. \quad (1)$$

Consistent with the previous studies (3, 10), the sign of $\nu_s(0)$ is dominated by the angle $\theta$ due to $b/l \ll 1$.

The scenario is different if the material is bent uniformly in the $x_1$ direction, though the deformation is still of similar mirror symmetry. In this situation, the struts parallel to the $x_1$ axis bend in the $x_1 - x_3$ plane, the sides of the star are twisted axially and bent out of the $x_1 - x_2$ plane, and the struts parallel to the $x_2$ axis sustain rigid body rotations in the $x_2 - x_3$ plane. The lattice material is curved, and the average curvatures $\kappa_{11}$ and $\kappa_{22}$ along the $x_1$ and $x_2$ directions can be calculated from the relative rotations between the unit normals at two opposite nodes of the unit cell. To classify the bending shapes, it is convenient to invoke the bending Poisson's ratio (BPR) $\nu_b(0) = -\kappa_{22}/\kappa_{11}$ (14), as $\nu_b(0) > 0$, $\nu_b(0) = 0$, and $\nu_b(0) < 0$ imply anticlastic, monoclastic, and synclastic curvatures, respectively. Our derivation provides (*SI Appendix*, S1B)

$$\nu_b(0) = \frac{\lambda - 1}{3\lambda + 1}\sin 2\theta, \quad (2)$$

where $\lambda$ is the ratio of torsion to bending stiffness of the beams, given by

$$\lambda = \frac{6b^2}{(1+\nu)h^2}\left(\frac{1}{3} - \frac{64b}{\pi^5 h}\sum_{n=1,3,5\ldots}^{\infty}\frac{1}{n^5}\tanh\frac{n\pi h}{2b}\right). \quad (3)$$

Accordingly, the sign of $\nu_b(0)$ is determined by both $h/b$ and $\theta$ .

As seen in Fig. 2*A*, for any Poisson's ratio $\nu$ of the beams, the value of $\lambda$ can be larger or less than 1 depending on the height-to-width ratio $h/b$ . Given the parameters $\theta$, $b$ and $l$, the value of $\nu_s(0)$ is fully determined, while the magnitude and sign of $\nu_b(0)$ are tunable by changing the height $h$ of the beams so that the lattice material can reveal saddle, cylinder, and dome shapes (Fig. 2*B*). Illustrated in Fig. 2*C* are the variations of $\nu_s(0)$ and $\nu_b(0)$ with $\theta$ for different values of the ratio $h/b$ at $b/l = 0.025$ and $\nu = 0.3$. The results agree well with FEMs for materials with more than $10 \times 10$ unit cells. We see that $\nu_s(0)$ is negative and positive respectively for $\theta < 71.6^\circ$ and $\theta > 71.6^\circ$, but the sign of $\nu_b(0)$ is convertible at any $\theta$ except $\theta = 90^{\circ}$, where $\nu_b(0) = 0$, when the ratio $h/b$ is changed across the value of about 0.56. For instance, at $\theta = 77^{\circ}$ and $h/b = 0.1$ both $\nu_s(0)$ and $\nu_b(0)$ are positive, the bending curvature is anticlastic. If we increase the height $h$ so that $h/b = 4.0$, $\nu_s(0)$ is invariant while $\nu_b(0)$ becomes negative, meaning that the material adopts a synclastic curvature. This breaks the belief that a lattice material with positive IPR bends into a saddle shape. To confirm, a rigid polymer specimen with $\theta = 77^{\circ}$, $h/b = 4$ and $b/l = 0.025$ was prepared by 3D printing. The specimen contracts transversely (positive IPR) if tensioned uniformly along the $x_1$ axis (Fig. 2*D*), and, indeed, displays a dome shape (negative BPR) if bent in the same direction (Fig. 2*D*). The cases when the material is loaded in other directions will be discussed later.

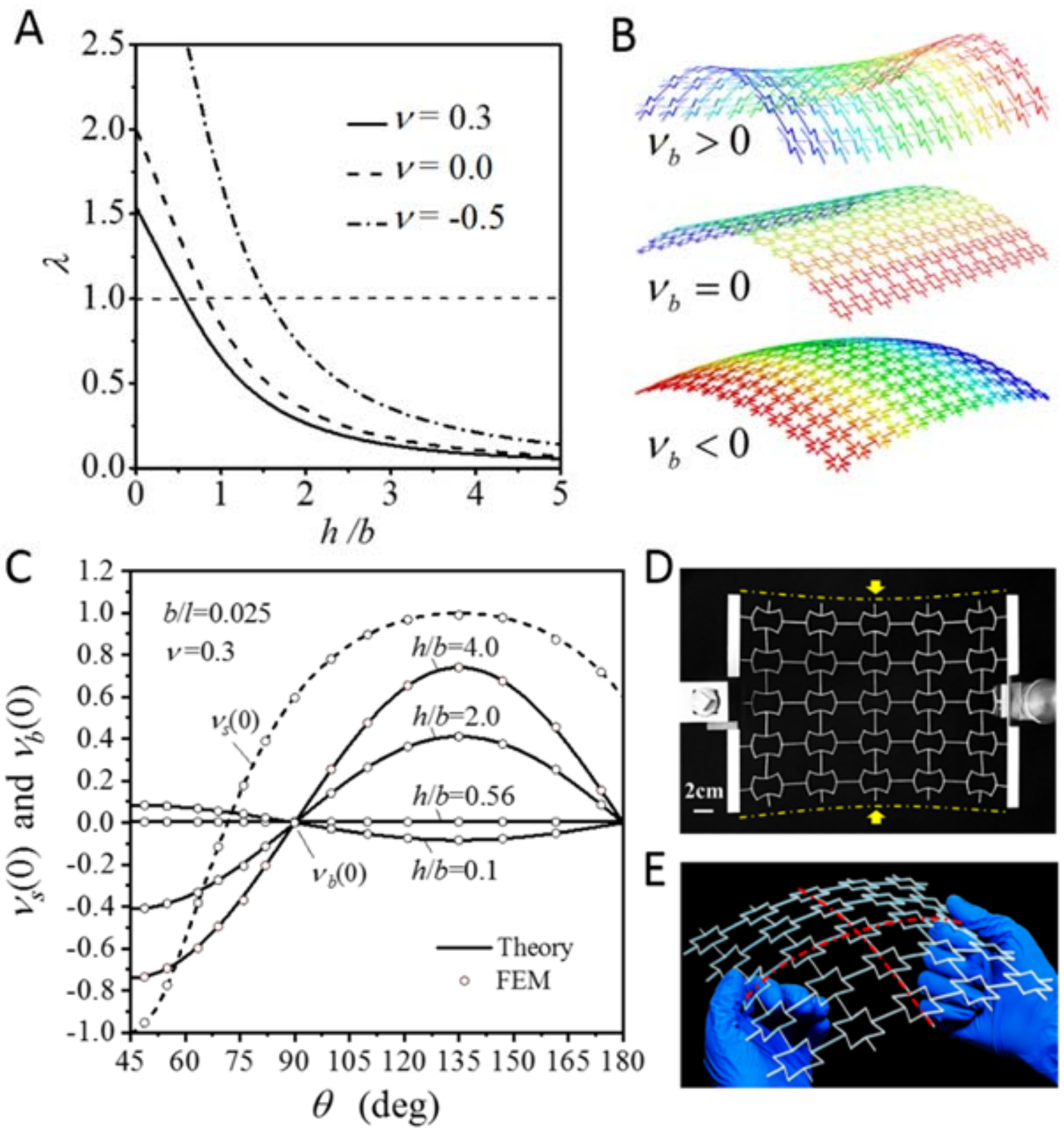

FIG. 2. (*A*) Dependence of the dimensionless parameter $\lambda$ on the ratio $h/b$. (*B*) Shape transition of the lattice material induced by changing $\lambda$. (*C*) Variations of the IPR $\nu_s(0)$ and BPR $\nu_b(0)$ with the angle $\theta$ when the tetragonal lattice material is tensioned and bent in the $x_1$ direction. The hollow circles represent the results of FEM simulations. (*D*) Tension test of a 3D printed tetragonal lattice material with $\theta = 77^{\circ}$, $b/l = 0.025$, and $h/b = 4.0$ in the $x_1$ direction. (*E*) Bending of the same 3D printed tetragonal lattice material in the $x_1$ direction.

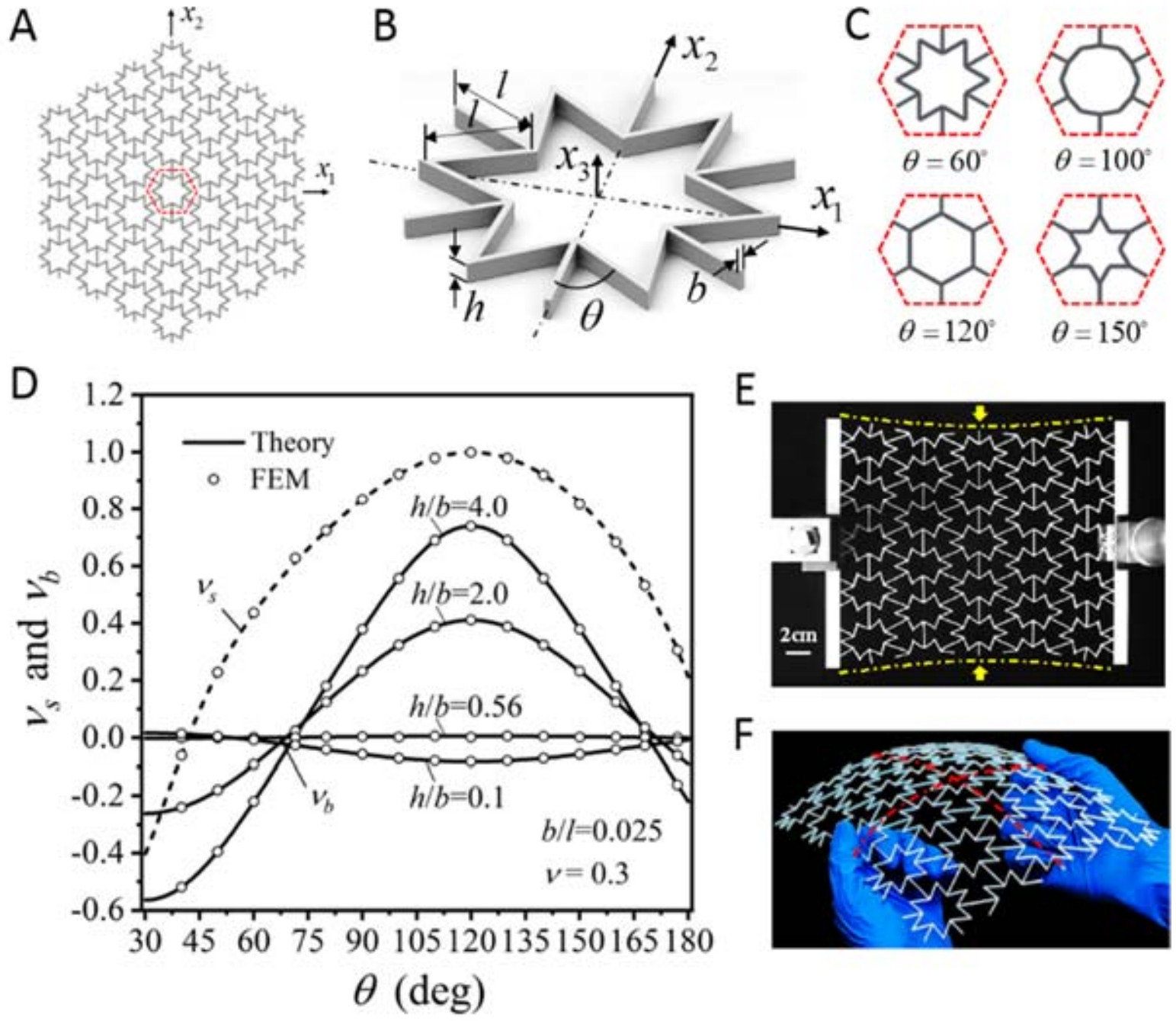

FIG. 3. (*A*) Structure of the hexagonal 2D lattice material. (*B*) Geometry of the 6-node hexagon unit cell. (*C*) Variants of the unit cell for different angle $\theta$. (*D*) Variations of the isotropic IPR $\nu_s$ and BPR $\nu_b$ with the angle $\theta$. The hollow circles represent the results of FEM simulations. (*E*) Tension test of a 3D printed hexagonal lattice material with $\theta = 55^\circ$, $b/l = 0.025$, and $h/b = 4.0$. (*F*) Bending of the same 3D printed hexagonal lattice material.

***Hexagonal 2D lattices***. Similar sign convertibility of BPR is also found in a hexagonal 2D lattice material with six-node star-shaped unit cells. Figs. 3*A*-*C* depict the structure of the material as well as the geometry and variants of a unit cell respectively. The length of the connecting struts is still assumed to be twice the side length of the star. The six-fold rotational symmetry dictates that the effective mechanical response of the material is isotropic in the $x_1 - x_2$ plane. We can arrive at

$$\begin{aligned} \nu_s &= \frac{\alpha_0(\theta) + \alpha_1(\theta)(b/l)^2 + \alpha_2(\theta)(b/l)^4}{\beta_0(\theta) + \beta_1(\theta)(b/l)^2 + \beta_2(\theta)(b/l)^4}, \\ \nu_b &= \frac{(\lambda - 1)\left[\eta_0(\theta) + \lambda\eta_1(\theta)\right]}{\xi_0(\theta) + \lambda\xi_1(\theta) + \lambda^2\xi_2(\theta)}, \end{aligned} \tag{4}$$

where the explicit expressions of $\alpha_k(\theta)$, $\beta_k(\theta)$, $\xi_k(\theta)$, and $\eta_k(\theta)$ ($k = 0$, 1, or 2) are given in *SI Appendix*, S2. Again, the sign of $\nu_s$ is dominated by the angle $\theta$ due to $b/l << 1$, and the sign of $\nu_b$ is determined by both $h/b$ and $\theta$. At the special case of $\theta = 120^\circ$, the material becomes a regular honeycomb (Fig. 3*C*), with $\nu_s = [4-(b/l)^2]/[4+3(b/l)^2]$ and $\nu_b = (1-\lambda)/(1+3\lambda)$. The former is well known (46), showing $\nu_s > 0$ for small $b/l$, while the latter indicates that $\nu_b$ can be positive or negative depending on $\lambda$. Variations of $\nu_s$ and $\nu_b$ with $\theta \in (30^\circ, 180^\circ)$ are plotted in Fig. 3*D* for $b/l = 0.025$ and $\nu = 0.3$, and the results are consistent with FEMs for large-sized samples with more than $10\times10$ unit cells. Apparently, $\nu_s$ is negative for $\theta < 37.4^\circ$ and positive for $\theta > 37.4^\circ$, and at any $\theta$ the sign of $\nu_b$ is convertible between positive and negative by changing the ratio $h/b$. Taking $\theta = 55^\circ$ as an example, $\nu_s$ is positive, while $\nu_b$ is negative at $h/b = 4$, so the material develops synclastic curvature. This is in contradiction with the common belief, but can be verified by using a 3D printed rigid polymer specimen with the same geometry. As shown in Figs. 3*E* and 3*F*, the specimen exhibits positive IPR but bends into a dome rather than a saddle shape.

***Cosserat continuum model.*** The above bending behavior cannot be described by a continuum model of classical elasticity, as the predicted IPR and BPR are identical (see *SI Appendix*, S3A). In reality, a unit cell in lattice materials is mapped to a small volume element in a continuum model. Classical elasticity treats the volume element as infinitesimal, characterizes its motion by three displacements, and neglects force couples acting on the surface. The situation is different from that in 2D lattices, where the in-plane dimension of unit cells may be significantly larger than the thickness, and force couples arising from bending and twisting transmitted through beams causes not only displacement but also rotation of a unit cell (47). To capture these complicated features, Cosserat elasticity (43, 44) turns out to be a proper choice as it endows a volume element with three rotations $\omega_i$ in addition to the displacements $u_i$ and allows for force couples by introducing couple stresses $m_{ij}$ in addition to the stresses $\sigma_{ij}$. The deformation is measured by strains $\varepsilon_{ij} = \partial_i u_j + e_{kji}\omega_k$ and torsions $k_{ij} = \partial_i \omega_j$, where $e_{ijk}$ .is the permutation symbol. In small deformations, the constitutive relations of a centrosymmetric solid read $\sigma_{ij} = A_{ijkl}\varepsilon_{kl}$ and $m_{ij} = B_{ijkl}k_{ij}$, with $A_{ijkl} = A_{jklij}$ and $B_{ijkl} = B_{jklij}$ being the fourth-order elasticity tensors. In general both $\sigma_{ij}$ and $m_{ij}$ are asymmetric, but satisfy the equilibrium equations $\partial_j \sigma_{ji} = 0$ and $\partial_j m_{ji} + e_{ikl}\sigma_{kl} = 0$.

We model the tetragonal lattice material as a Cosserat plate with the same thickness and symmetry. In this case each of $A_{ijkl}$ and $B_{ijkl}$ has 9 independent components. External loads applied to the boundary are characterized by membrane forces $N_{\alpha\beta}$ and moments $M_{\alpha\beta}$ defined by

$$\begin{aligned} N_{\alpha\beta} &= \int_{-h/2}^{h/2} \sigma_{\alpha\beta} \mathrm{d}x_3, \\ M_{\alpha\beta} &= \int_{-h/2}^{h/2} (m_{\alpha\beta} + e_{3\lambda\beta}\sigma_{\alpha\lambda} x_3) \mathrm{d}x_3, \end{aligned} \tag{5}$$

where a Greek subscript is equal to 1 or 2. Let $x_1'$ and $x_2'$ be two orthogonal in-plane coordinates, with $x_1'$ making an angle $\phi$ with the $x_1$ axis. If the plate is uniformly tensioned and bent in the $x_1'$ direction by $N_{11}' = N_0$ and $M_{12}' = M_0$, we can derive that (*SI Appendix*, S3B)

$$\begin{aligned} \varepsilon_{11}'^{0} &= N_0 (Q_{11} + Q_{12}\cos 4\phi), \\ \varepsilon_{22}'^{0} &= N_0 (Q_{22} - Q_{12}\cos 4\phi), \\ \kappa_{11}' &= -M_0 (D_{11} + D_{12}\cos 4\phi), \\ \kappa_{22}' &= -M_0 (D_{22} - D_{12}\cos 4\phi), \end{aligned} \tag{6}$$

in which $Q_{\alpha\beta}$ and $D_{\alpha\beta}$ are the combinations of $A_{ijkl}, B_{ijkl}$ and $h$. For equivalency, the results must be identical to those of the square unit cell shown in Fig. 1*B* when it is subjected to the same loads. This, at $\phi = 0^\circ$ and $\phi = 45^\circ$, leads to 6 independent equations which allow us to determine $Q_{\alpha\beta}$ and $D_{\alpha\beta}$ from the lattice material (*SI Appendix*, S3B). The IPR $\nu_s(\phi) = -\varepsilon_{22}'^{0} / \varepsilon_{11}'^{0}$ and BPR $\nu_b(\phi) = -\kappa_{22}' / \kappa_{11}'$ are finally obtained as

$$\begin{aligned} \nu_s(\phi) &= \frac{5f^2(\theta) - g(\theta,\phi) - 2\left[f^2(\theta)\cos^2 2\phi - 1\right](b/l)^2}{3f^2(\theta) - g(\theta,\phi) + 2\left[f^2(\theta)\sin^2 2\phi - 3\right](b/l)^2}, \\ \nu_b(\phi) &= \frac{(\lambda - 1)\left[f^2(\theta)\cos^2 2\phi - 1\right]}{3\lambda + 1 - (\lambda - 1) f^2(\theta)\sin^2 2\phi}, \end{aligned} \tag{7}$$

with $f(\theta) = \sin\theta + \cos\theta$ and $g(\theta,\phi) = 8 + 4\left[3 - 3f(\theta) + 2f^2(\theta)\right]\sin^2 2\phi$. Similarly, the hexagonal lattice material can be modeled as a transversely isotropic Cosserat plate with 16 independent elastic constants. Note that the relations in Eq. 6 still hold, except $Q_{12} = D_{12} = 0$. Thus, the IPR $\nu_s(\phi) = -Q_{22}/Q_{11}$ and BPR $\nu_b(\phi) = D_{22}/D_{11}$ are isotropic and must be equal to the results given in Eq. **4**.

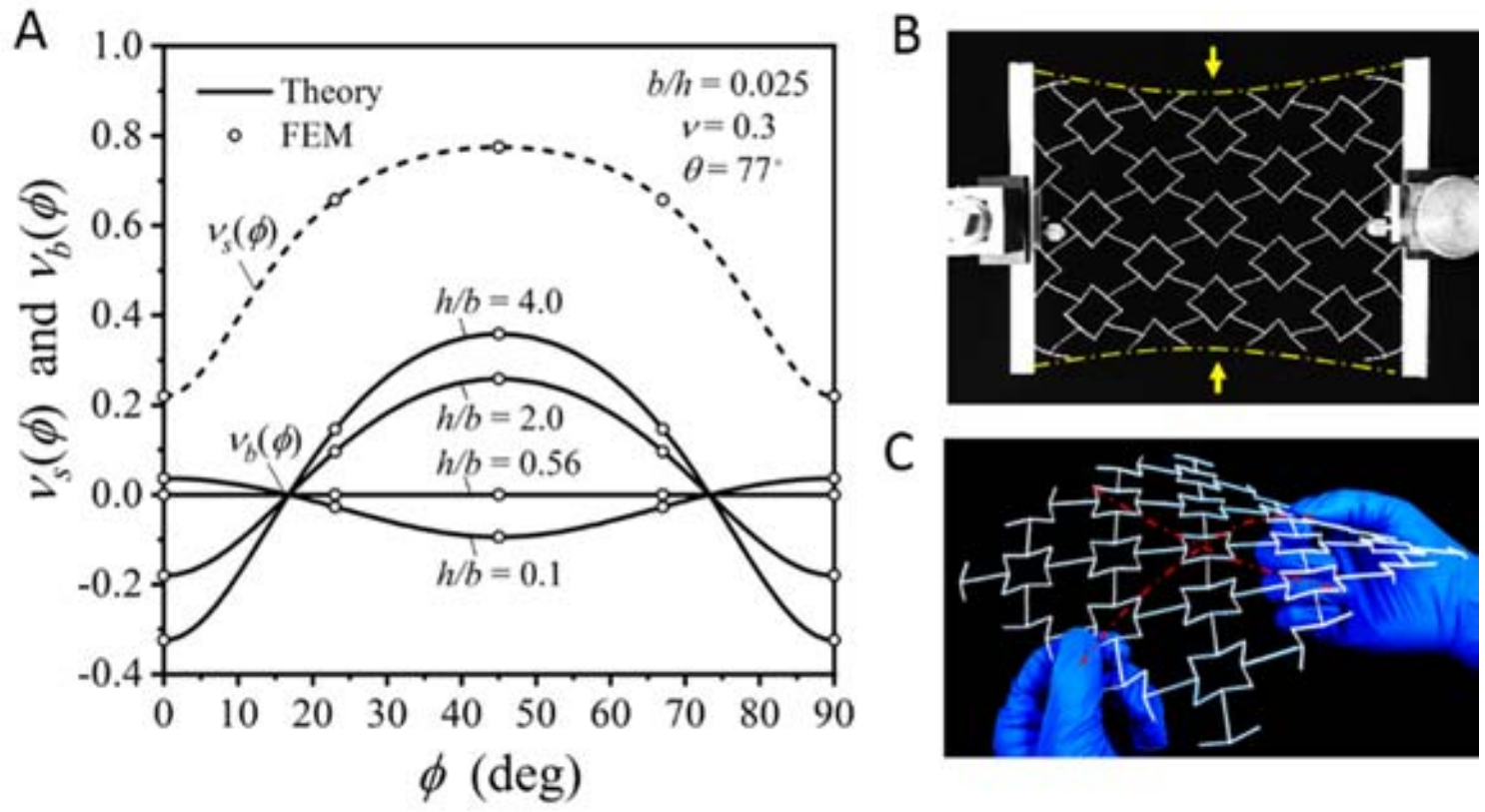

FIG. 4. (*A*) Dependence of the IPR $\nu_s(\phi)$ and BPR $\nu_b(\phi)$ of the tetragonal 2D lattice material on the loading direction $\phi$. The hollow circles are the results of FEM simulations. (*B*) Tension test of a 3D printed tetragonal lattice material with $\theta = 77^\circ$, $b/l = 0.025$, and $h/b = 4.0$ in the direction of $\phi = 45^\circ$. (*C*) Bending of the same specimen in the direction of $\phi = 45^\circ$.

Eq. **7** quantifies the direction dependence of the IPR and BPR of the tetragonal lattice material. In view of the four-fold symmetry, we only need to discuss the case of $\phi \in [0^\circ, 90^\circ]$. Fig. 4*A* demonstrates the variations of $\nu_s(\phi)$ and $\nu_b(\phi)$ with $\phi$ for $b/l = 0.025$, $\nu = 0.3$, and $\theta = 77^\circ$, where the results of FEM are also provided for validation. It is visible that $\nu_s(\phi)$ is always positive, while $\nu_b(\phi)$ exhibits changeable signs depending not only on the angle $\phi$ but also on the ratio $h/b$. At $h/b = 0.56$ ($\lambda = 1$), $\nu_b(\phi)$ vanishes for any $\phi$, meaning that the lattice material adopts a cylinder shape when bent in an arbitrary direction. For $h/b > 0.56$, $\nu_b(\phi)$ is positive at $\phi \in (16.8^\circ, 73.2^\circ)$ and negative elsewhere; for $h/b < 0.56$, $\nu_b(\phi)$ becomes negative at $\phi \in (16.8^\circ, 73.2^\circ)$ and positive elsewhere. The former is consistent with the experiment described in Figs. 2*D* and 2*E*, and the latter can be examined by testing the same 3D printed rigid polymer specimen along the direction of $\phi = 45^\circ$. As seen in Figs. 4*B* and *C*, the specimen contracts transversely but bends into a saddle shape. This confirms the theoretical prediction.

## 3. Conclusion

In summary, pure bending of architected 2D beam lattices with star-shaped unit cells has been examined. The bending shape was shown to be dependent on the re-entrant angle of the stars and the cross-sectional ratio of the beams. It is well known that change in the re-entrant angle leads to positive, zero, or negative IPR of the material. However, an unexpected finding is that at any nonzero IPR there exists a transition between anticlastic and synclastic curvatures controlled by the beam's cross-sectional ratio. This defies the common belief that saddle- and dome-like bending shape of a 2D lattice is dictated by the sign of its IPR. Classical elasticity does not have sufficient freedom to capture the complicated behavior, and for an appropriate macroscopic description one has to resort to some higher-order continuum models such as Cosserat elasticity. Our work highlights the importance of competitive interactions between axial torsion and out-of-plane bending of the beams, and the results may generate a number of theoretical and practical consequences in the study of more complex 2D lattice systems.

## 4. Materials and Methods

***Experimental***. The specimens used in our experiments are rigid polymer (VeroWhite) lattice structures prepared by 3D printing from designed CAD models by using an Objet260 Connex 3D printer (Stratasys Ltd., USA). The geometric parameters of the tetragonal lattice are $\theta = 77^\circ$, $l = 12$ mm, $b = 0.3$ mm, and $h = 1.2$ mm, while those of the hexagonal lattice are $\theta = 55^\circ$, $l = 12$ mm, $b = 0.3$ mm, and $h = 1.2$ mm. Each specimen has two much thicker and stiffer frames on two opposite sides, so that uniform displacements can be applied to the boundary intension tests. The frames then are cut in bending tests. Tension test is conducted via In Situ Microscope with Microtest (JP-Nikon TiS; GB-Microtest 5000W). As uniform bending with constant moment is hard to realize, we bend a specimen with hands.

***Numerical Model***. The commercial software ABAQUS/Standard is employed for the FEM simulations, where the lattice structure is meshed by B33 Euler-Bernoulli beam element. The beams in the lattice materials are in rigid connection, with length $l = 12$ mm, width $b = 0.3$ mm, Yong's modulus $E = 1.0$ GPa, and Poisson's ratio $\nu = 0.3$. The half re-entrant angle $\theta$ of the stars and the thickness $h$ of the beams are allowed to change. In the simulations of uniform tension and bending, displacements and rotations proportional to the coordinates are prescribed at each nodes on the boundary of the lattice materials respectively. Each sample for computation contains at least $10 \times 10$ unit cells so as to eliminate boundary effects (*SI Appendix*, Fig. S9).

**Acknowledgments**. This work was supported by the National Natural Science Foundation of China (Nos. 12072337, 12025206).